\documentclass[12pt]{article}

\usepackage{graphics}
\usepackage{color}
\usepackage{amssymb,amsmath}
\usepackage{slashed}
\usepackage{epstopdf}
\usepackage{booktabs}
\usepackage{mathrsfs}
\usepackage[export]{adjustbox}
\usepackage{epsfig}
\usepackage{graphicx,color}

\newcommand{\bea}{\begin{eqnarray}}
\newcommand{\eea}{\end{eqnarray}}

\numberwithin{equation}{section}

\begin{document}
\begin{titlepage}
%\begin{flushright}
%
%\end{flushright}
%
\vspace*{10mm}
\begin{center}
\baselineskip 25pt 
{\Large\bf
%%%%%%%%%%%%%%%%%%%%%%%%%%%%%%%%%%%%%%%%%%%%%%%%%%%
Inflection-point $B-L$ Higgs Inflation
%%%%%%%%%%%%%%%%%%%%%%%%%%%%%%%%%%%%%%%%%%%%%%%%%%%
}
\end{center}
\vspace{5mm}
\begin{center}
{\large
Nobuchika Okada\footnote{okadan@ua.edu}
and
Digesh Raut\footnote{draut@crimson.ua.edu}
}
\end{center}
\vspace{2mm}

\begin{center}
{\it
Department of Physics and Astronomy, University of Alabama, \\
Tuscaloosa, Alabama 35487, USA
}
\end{center}
\vspace{0.5cm}
%%%%%%%%%%%%%%%%%%%%%%
\begin{abstract}
%%%%%%%%%%%%%%%%%%%%%%
Inflection-point inflation is an interesting possibility to realize a successful slow-roll inflation
when inflation is driven by a single scalar field with its initial value below the Planck mass ($\phi_I \lesssim M_{Pl}$). 
In order for a renormalization group (RG) improved effective $\lambda  \phi^4$ potential
to develop an inflection-point, the quartic coupling $\lambda(\phi)$  must exhibit a minimum
with an almost vanishing value in its RG evolution, namely $\lambda(\phi_I) \simeq 0$ and $\beta_{\lambda}(\phi_I) \simeq 0$, where $\beta_{\lambda}$ is the beta-function of the quartic coupling. 
As an example, we consider the minimal gauged $B-L$ extended Standard Model at the TeV scale, where we identify the $B-L$ Higgs field as the inflaton field. 
For a successful 
inflection-point inflation, which is consistent with the current cosmological observations, the mass ratios among the $Z^{\prime}$ gauge boson, the right-handed neutrinos
and the $B-L$ Higgs boson are fixed. 
Our scenario can be tested in the future collider experiments such as the High-Luminosity LHC and the SHiP experiments. 
In addition, the inflection-point inflation provides a unique prediction for the running of the spectral index $\alpha \simeq  - 2.7 \times 10^{-3}\left(\frac{60}{N}\right)^2$ ($N$ is the $e$-folding number), which can be tested in the near future. 

%%%%%%%%%%%%%%%%%%%%%%%
\end{abstract}
\end{titlepage}

%%%%%%%%%%%%%%%%%%%%%%%%%%%%%%%%%
\section{Introduction}
%%%%%%%%%%%%%%%%%%%%%%%%%%%%%%%%%
Current understanding of the cosmic origin and the evolution is that
our 
	universe went through a period of rapid accelerated expansion at the beginning, which is known as inflation. 
Inflation \cite{inflation1, inflation2, chaotic_inflation, inflation3} solves several serious problems of the Standard Big Bang 
 Cosmology,
 	such as the horizon, flatness and monopole problems.
More importantly, the primordial density fluctuations generated during inflation seed the formation of large scale structure of the universe we see today. 
In a simple inflationary scenario known as slow-roll inflation, 
    inflation is driven by a single scalar field (inflaton) when it slowly rolls down to its potential. 
During the slow-roll, the energy density of the universe is dominated by the inflaton potential energy,
which drives accelerated expansion of the universe. 
After the end of inflation, 
	the inflaton decays to Standard Model (SM) particles to reheat 
the universe, and the Standard Big Bang Cosmology begins.

The slow-roll inflation requires the inflaton potential to be sufficiently flat in the inflationary epoch. 
In chaotic inflation, such a flat potential is realized by taking initial inflaton value to be of the trans-Planckian scale.  
From the field theoretical point of view, effective operators suppressed by the Planck mass ($M_{Pl}= 1.22\times10^{19}$ GeV)
	could significantly contribute to the inflaton potential and hence the inflationary predictions.
For this reason, it may be more appealing to consider the small-field inflation (SFI) scenario, where the initial inflation value is smaller than the Planck mass. 
Hybrid inflation, which is realized with multiple scalar fields, is a well known example of the SFI \cite{HybridInflation}.
If one considers the SFI driven by a single scalar field, the so-called inflection-point inflation is an interesting possibility \cite{SF_Inf, InflectionPoint, InflectionPoint2}. 
In this scenario the potential has an inflection-point, and hence slow-roll inflation can be realized if the initial inflaton value is taken in the immediate vicinity of the inflection-point.

An inflation scenario is more compelling if the inflaton field plays another important role in particle physics. 
It is then interesting to identify the inflaton (scalar) field with a Higgs field in a general Higgs model, which plays a crucial role to spontaneously break a gauge symmetry of the model. 
For example, in references \cite{Higgs_inflation1,Higgs_inflation2,Higgs_inflation3},  the SM Higgs field is identified with the inflaton field. 
In this paper, we investigate the inflection-point inflation in a general Higgs model, where the inflaton field is identified with the Higgs field and has both gauge and Yukawa interactions, just like the SM.

To realize the inflection-point in a Higgs/inflaton potential, we consider a Renormalization-Group (RG) improved effective Higgs/inflaton potential. 
During the inflation, we assume that inflaton value is much larger than its Vacuum Expectation Value (VEV) at the potential minimum, so that the inflaton potential is dominated by its quartic term. 
If the RG running of the inflaton quartic coupling first decreases towards high energy and then increases, inflection-point is realized in the vicinity of the minimum point of the running quartic coupling, where both the quartic coupling and its beta-function become vanishingly small \cite{InflectionPoint, InflectionPoint2}\footnote{In the context of the $\lambda \phi^4$ inflation with a non-minimal gravitational coupling \cite{NonMinimalUpdate}, similar conditions have been derived to ensure the stability of the inflaton potential \cite{RunningInflation1}.}. 
Interestingly, these boundary conditions lead to correlations between the very high energy physics of inflation and the low energy particle phenomenology.

For simplicity, let us consider a Higgs model with its RG improved effective potential given by
\bea
   V(\phi) = \frac{1}{4} \lambda(\phi) \;  \phi^4, 
\eea
    where $\phi$ denotes the inflaton field, $\lambda(\phi)$ is the running quartic coupling, and we have neglected the mass term assuming the initial inflaton value is much larger than the mass term. 
The running coupling coupling satisfies the (one-loop) RG equation of the form,  
\bea 
  16 \pi^2 \frac{d \lambda}{d \ln \mu} \simeq  C_g \; g^4  -  C_Y \; Y^4 ,
  \label{RGgeneral} 
\eea  
where $g$ and $Y$ are the gauge and Yukawa couplings, respectively, 
    and $C_g$ and $C_Y$ are positive coefficients 
whose actual values are calculable once the particle contents of the model are defined.  
In Eq.~(\ref{RGgeneral}) we have neglected terms proportional to $\lambda$ ($\lambda^2$ term and the anomalous dimension term) because the SFI requires the quartic coupling $\lambda \propto g^6$, as will be shown later.  
Hence the quantum corrections to the effective Higgs potential are dominated by the gauge and Yukawa interactions.   
Realization of the inflection-point requires a vanishingly small beta-function at the initial inflaton value, namely $C_g \;  g - C_Y \;  Y = 0 $.  
This condition leads to a relation between $g$ and $Y$, or in other words, 
    the mass ratio of gauge boson to the fermion in the Higgs model is fixed. 
Since the Higgs quartic coupling at low energy is evaluated by solving the RG equation, 
   the resultant Higgs mass also has a unique relation to the gauge and the fermion masses.

As a concrete example of such a model, in this paper we consider the minimal gauged $B-L$ (baryon number minus lepton number) extension of the SM,
   where the global $B-L$ symmetry in the SM is gauged.  
The model has three right-handed neutrinos and the $B-L$ Higgs field (identified with inflaton), which
   are introduced for the cancellation of the gauge and gravitational anomaly and the $B-L$ gauge symmetry breaking, respectively.  
Associated with the $B-L$ gauge symmetry breaking, the $B-L$ gauge boson 
   and the right-handed neutrinos acquire their masses. 
With the Majorana masses for the right-handed neutrinos, 
   the seesaw mechanism \cite{Seesaw}, which naturally realizes the light neutrino mass generation, is automatically implemented in this model.

The paper is organized as follows. 
In the next section, we give a brief review of the slow-roll inflation. 
In Sec.~3, we present the inflationary predictions for the scenario, where the inflaton potential exhibits an inflection-point-like behavior during the slow-roll. 
In Sec.~4, we consider the minimal gauged $B-L$ extension of the SM, where the $B-L$ Higgs field is identified with the inflaton field. 
To realize the inflection-point in a Higgs/inflaton potential, we consider the RG improved effective Higgs/inflaton potential. 
In Sec.~5, we consider the constraints on the model parameters from the Big Bang Nucleosynthesis and the current collider experiments. 
We also discuss the prospects of testing the scenario in the future collider experiments,  such as the High-Luminosity LHC and SHiP experiments. 
Sec.~6 is devoted to conclusions.

Before moving on to the next section we comment on the differences between this work and the previous work in Ref.~\cite{InflectionPoint2}. 
Although the parameterization of our inflaton potential is slightly different, our discussions in Sec.~3 and Sec.~4 are well overlapping with those in Ref.~\cite{InflectionPoint2}. 
However, the authors in Ref.~\cite{InflectionPoint2} mainly focus on the the inflection-point inflation with a large inflaton value beyond the Planck scale in the presence of non-minimal coupling while we focus on the SFI, so that our parameter regions are very different. 
More importantly our motivation for this work is to consider a complementarity between the inflection-point inflation and the new physics search at low energies as we will discuss in Sec.~5.  

%%%%%%%%%%%%%%%%%%%%%%%%%%%%%%%%%%%%%%
\section{Brief Review of Slow-roll Inflation}
%%%%%%%%%%%%%%%%%%%%%%%%%%%%%%%%%%%%%%

The inflationary slow-roll parameters for the inflaton field ($\phi$) are expressed as 
\bea
\epsilon(\phi)=\frac{ M_{P}^2}{2} \left(\frac{V'}{V}\right)^2, \; \; 
\eta(\phi)=
M_{P}^2\left(\frac{V''}{V }\right), \;\;
\zeta^2{(\phi)} = M_{P}^4  \frac{V^{\prime}V^{\prime\prime\prime}}{V^2}, 
 \label{SRCond}
\eea
where $M_{P}= M_{Pl}/\sqrt{8 \pi} = 2.43\times 10^{18}$ GeV is the reduced Planck mass, $V$ is the inflation potential, and the prime denotes the derivative with respect to $\phi$.  
The amplitude of the curvature perturbation $\Delta^2_{\mathcal{R}}$ is given by 
\begin{equation} 
\Delta_{\mathcal{R}}^2 = \frac{1}{24 \pi^2}\frac{1}{M_P^4}\left. \frac{V}{ \epsilon } \right|_{k_0},
 \label{PSpec}
\end{equation}
  which should satisfy $\Delta_\mathcal{R}^2= 2.195 \times10^{-9}$
  from the Planck 2015 results \cite{Planck2015} with the pivot scale chosen at $k_0 = 0.002$ Mpc$^{-1}$.
The number of e-folds is defined as
\begin{eqnarray}
N=\frac{1}{M_{P}^2}\int_{\phi_E}^{\phi_I}\frac{V }{V^\prime} d\phi  ,
 \label{EFold}
\end{eqnarray} 
where $\phi_I$ is the inflaton value at a horizon exit corresponding to the scale $k_0$, 
  and $\phi_E$ is the inflaton value at the end of inflation, 
  which is defined by $\epsilon(\phi_E)=1$.
The value of $N$ depends logarithmically on the energy scale during inflation 
  as well as on the reheating temperature, and it is typically taken to be 50--60.

The slow-roll approximation is valid as long as the conditions 
   $\epsilon \ll 1$, $|\eta| \ll 1$ , and $\zeta^2\ll1$ hold. 
In this case, the inflationary predictions are given by
\bea
n_s = 1-6\epsilon+2\eta, \; \; 
r = 16 \epsilon , \;\;
\alpha = 16 \epsilon \eta -24 \epsilon^2-2 \zeta^2, 
 \label{IPred}
\eea 
where $n_{s}$ and $r$ and $\alpha \equiv \frac{\mathrm{d}n_s}{d ln k}$ are the scalar spectral index, the tensor-to-scalar ratio and the running of the spectral index, respectively, which are evaluated at $\phi = \phi_I$.  
The Planck 2015 results \cite{Planck2015} set an upper bound on the tensor-to-scalar ratio as $r \lesssim 0.11$,  
         while the best fit value for the spectral index ($n_s$) and the running of spectral index ($\alpha$) are $0.9655 \pm 0.0062$ and $−0.0057\pm 0.0071$, respectively, at $68 \%$ CL.

 %%%%%%%%%%%%%%%%%%%%%%%%%%%%%%%%%%%%%%
\section{Inflection-point Inflation}
%%%%%%%%%%%%%%%%%%%%%%%%%%%%%%%%%%%%%% 
In the SFI scenario, to realize the slow-roll inflation the inflaton potential must exhibit an inflection-point-like behavior, where the potential is very flat\footnote{For successful inflation scenario it is not necessary for the potential to realize an {\it exact} inflection-point. 
We only require the inflaton potential to exhibit a behavior of almost an inflection-point.}.
The initial inflaton value is set in the very flat region $\phi_I = M$. 
We consider the following expansion of an inflaton potential around the $\phi =M$ given by 
\bea
V(\phi)\simeq V0 +V1 (\phi-M)+\frac{V2}{2} (\phi-M)^2+\frac{V3}{6} (\phi-M)^3, 
\label{PExp}
\eea
where $V0$ is constant and $V1$, $V2$ and $V3$ are the first, second and third derivatives of the inflaton potential evaluated at $\phi=M$. 
To realize a very flat potential with an inflection-point-like behavior, we require $V1$ and $V2$ to be vanishingly small.  
From Eqs.~(\ref{SRCond}) and (\ref{PExp}), the slow-roll parameters are then given by 
\bea
\epsilon(M) \simeq \frac{M_{P}^2}{2}\Big(\frac{V1}{V0}\Big)^2, \;\;
\eta(M) \simeq M_{P}^2\Big(\frac{V2}{V0}\Big), \;\;
\zeta^2{(M)} = M_{P}^4  \frac{V1V3}{V0^2}, 
\label{IPa}
\eea
where we have used the approximation $V(M)\simeq V0$. 
Similarly, the power-spectrum $\Delta_{\mathcal{R}}^2$ is expressed as
\bea
\Delta_{\mathcal{R}}^2 \simeq \frac{1}{12\pi^2}\frac{1}{M_P^6}\frac{V0^3}{V1^2}.
\label{CV1} 
\eea 
Using the observational constraint, $\Delta_{\mathcal{R}}^2= 2.195 \times  10^{-9}$, and a fixed $n_s$ value, we obtain 
\bea
\frac{V1}{M^3}&\simeq& 1961\left(\frac{M}{M_P}\right)^3\left(\frac{V0}{M^4}\right)^{3/2}, \nonumber \\
\frac{V2}{M^2}&\simeq& -1.725\times 10^{-2}\Big(\frac{1-n_s}{1-0.9655}\Big)\Big(\frac{M}{M_P}\Big)^2\left(\frac{V0}{M^4}\right), 
\label{FEq-V1V2}
\eea
where we have used $V(M)\simeq V0$ and $\epsilon(M) \ll \eta(M)$ as we we see later. 
For the remainder of the analysis we set $n_{s}=0.9655$ (the center value from the Planck 2015 results \cite{Planck2015}). 
The inflaton value 
	at the end of inflation is parameterized as $\phi_E/M = 1- \delta_E $, where $\delta_E <1$. 
We define the inflaton value ($\phi_E$) at the end of inflation by $\epsilon(\phi_E)=1$.

Using Eq.~(\ref{EFold}), the $e$-folding number ($N$) is given by
\bea
N=\frac{2 V0}{M_{P}^2\sqrt{-V2^2+2 V1 V3} }\arctan \left(\frac{V2 + V3(\phi-M) }{\sqrt{-V2^2+2 V1 V3}}\right)\Big|_{\phi=M- M \delta_E}^{\phi=M}. 
\label{CV3} 
\eea
The inflection-point-like behavior requires $V1 \simeq 0$, $V2 \simeq 0$, and $V3 \neq 0$,
so we approximate $-V2^2+2 V1 V3 \simeq 2 V1 V3$. 
We will confirm later that this is a good approximation. 
We will also show later that $V2,  \sqrt{2 V1 V3}\ll V3 M \delta_E $. 
Hence the $e$-folding number is approximated as
\bea
N\simeq 2\chi \arctan \left[\frac{V3 M \delta_E }{\sqrt{2 V1 V3}}\right] \simeq \pi \chi, 
\label{CV4} 
\eea
where $\chi = \frac{V0}{M_{P}^2\sqrt{2 V1 V3}}$. 
Using Eq.~(\ref{FEq-V1V2}), $V3$ is then given by 
\bea
\frac{V3}{M} \simeq 6.989 \times 10^{-7}\Big(\frac{60}{N}\Big)^2 \Big(\frac{V0^{1/2}}{M M_P}\Big) . 
\label{FEq-V3} 
\eea 
From Eqs.~(\ref{FEq-V1V2}) and (\ref{FEq-V3}), we obtain $2 V1 V3 \simeq 9.2 (60/N) V2^2 $. 
Hence, for $N\simeq 60$ $(-V2^2+2 V1 V3)^{1/2} \simeq (2 V1V3)^{1/2}$ is a good approximation.

Using Eqs.~(\ref{IPred}), (\ref{FEq-V1V2}) and (\ref{FEq-V3}), we now express all the inflationary predictions in terms $V0$, $M$ and $N$. 
From Eqs.~(\ref{IPa}) and (\ref{FEq-V1V2}), tensor-to-scalar ratio ($r$) is given by
\bea 
r=3.077\times 10^7\Big( \frac{V0}{M_P^4}\Big). 
\label{FEq-r}
\eea 
The running of the spectral index ($\alpha$) is given by 
\bea
\alpha \simeq - 2\zeta^2(M) = - \; 2.742 \times 10^{-3}\left(\frac{60}{N}\right)^2, 
\label{FEq-alpha}
\eea
which is independent of $V0$ and $M$. 
This prediction is consistent with the  current experimental bound, $\alpha=−0.0057\pm 0.0071$ \cite{Planck2015}. 
Precision measurement of the running of the spectral index in future experiments can reduce the error to $\pm0.002$ \cite{RunningSpectral}. 
Hence, the prediction can be tested in the future.

%%%%%%%%%%%%%%%%%%%%%%%%%%%%%%%%%%%%%%
\section{The Inflection-point-like $B-L$ Higgs Inflation}
%%%%%%%%%%%%%%%%%%%%%%%%%%%%%%%%%%%%%%
\begin{table}[h]
\begin{center}
\begin{tabular}{c|ccc|c}
            & SU(3)$_c$ & SU(2)$_L$ & U(1)$_Y$ & U(1)$_{B-L}$  \\
\hline
$ q_L^i $    & {\bf 3}   & {\bf 2}& $+1/6$ & $+1/3$  \\ 
$ u_R^i $    & {\bf 3} & {\bf 1}& $+2/3$ & $+1/3$  \\ 
$ d_R^i $    & {\bf 3} & {\bf 1}& $-1/3$ & $+1/3$  \\ 
\hline
$ \ell^i_L$    & {\bf 1} & {\bf 2}& $-1/2$ & $-1$  \\ 
$ N_R^i$   & {\bf 1} & {\bf 1}& $ 0$   & $-1$  \\ 
$ e_R^i  $   & {\bf 1} & {\bf 1}& $-1$   & $-1$  \\ 
\hline 
$ H$         & {\bf 1} & {\bf 2}& $-1/2$  &  $ 0$  \\ 
$ \varphi$      & {\bf 1} & {\bf 1}& $  0$  &  $+2$  \\ 
\end{tabular}
\end{center}
\caption{
Particle contents of the minimal $B-L$ model. 
In addition to the SM particle contents, the right-handed neutrino $N_R^i$ 
 ($i=1,2,3$ denotes the generation index)  and a complex scalar $\varphi$ are introduced. 
}
\end{table}
%%%%%%%%%%%%%%%%%%%%%%%%%%%%%%%%%%%%%%%%%%%%%%%
As a simple example of the inflection-point  Higgs inflation, in this section we consider the minimal $B-L$ extension of the SM. 
Here, anomaly-free U(1)$_{B-L}$ gauge symmetry is introduced 
   along with a scaler field $\varphi$ ($B-L$ Higgs) and three right-handed neutrinos ($N_R^i$) which are necessary for the cancellation of all the anomalies. 
The particle contents of the model are listed in Table~1. 
For the right-handed neutrinos we add Majorana Yukawa interaction terms, 
\bea  
   {\cal L} \supset  - \frac{1}{2} \sum_{i=1}^{3} Y_i  \varphi  \overline{N_{R}^{i~C}} N^i_{R}  +{\rm h.c.}
\eea  
Associated with the gauge symmetry breaking, all the new particles, $B-L$ gauge boson ($Z^\prime$), the right-handed neutrinos ($N_R$) and the $B-L$ Higgs acquire their masses as follows:
\bea 
m_{Z^\prime}= 2 \; g \; v_{BL},  \; \; 
m_{N^i}= \frac{1}{\sqrt{2}} \; Y_i \; v_{BL}, \; \; 
m_{\phi} = \sqrt{2 \lambda} \;  v_{BL}, 
\label{masses}
\eea
where $v_{BL}= \sqrt{2} \langle \varphi\rangle$ is the VEV of the $B-L$ Higgs field.

For simplicity, we consider a scenario where the $B-L$ Higgs sector is very weakly coupled to the SM Higgs doublet. 
Hence, for the inflationary analysis, the $B-L$ Higgs/inflaton sector can be treated independently.  
The tree level potential for the $B-L$ Higgs field is given by 
\bea
  V_{tree}= \lambda_{tree} \left( \varphi^\dagger \varphi - \frac{v_{BL}}{2}  \right)^2.   
\eea
We redefine the $B-L$ Higgs field as $\varphi = (\phi+v)/\sqrt{2}$ in the unitary gauge, where $\phi=\sqrt{2}\Re[\varphi]$ is the physical $B-L$ Higgs boson, identified as the inflaton. 
For the inflationary analysis, we consider $v_{BL} \ll \phi_I$, so that the inflaton potential is approximately given by $V_{tree}= (1/4)  \lambda_{tree} \phi^4$, at the tree-level. 
In our analysis, we employ the RG improved effective potential given by  
\bea
V(\phi) = \frac{1}{4} \lambda (\phi)\;\phi^4, 
\label{VEff}
\eea
where $\lambda (\phi)$ is the solution to the following RG equations:
\bea
\phi  \frac{d g}{d \phi} &=& \frac{1}{16 \pi^2}12 g^3,         \nonumber\\
 \phi \frac{d Y_i}{d \phi}   &=& \frac{1}{16 \pi^2}\left(Y_i^3+\frac{1}{2}Y_i \sum_j Y_j^2-6 g^2 Y_i\right),   \nonumber\\
 \phi \frac{d \lambda}{d \phi}  &=& \beta_\lambda. 
  \label{RGEs}
\eea
Here, the beta-function of the inflaton quartic coupling ($\beta_\lambda$) is expressed as
\bea
\beta_\lambda = \frac{1}{16 \pi^2} \left(20 \lambda^2- 48\lambda  g^2+2 \lambda\sum_i Y_i^2  +96 g^4 - \sum_i Y_i^4\right).
\label{BGen}
\eea

We now turn to the inflationary analysis of the RG improved inflaton potential. 
At first, we simplify the model, and consider the degenerate mass spectrum for the right-handed neutrinos, $Y \equiv Y_1 = Y_2=Y_3$. 
Thus the beta-function of the quartic coupling is 
\bea
\beta_\lambda=\frac{1}{16 \pi^2}\left(20 \lambda^2- (48 g^2 -6 Y^2) \lambda +96 g^4 - 3 Y^4\right). 
\label{BDeg}
\eea

Now we express the coefficients in the expansion of Eq.~(\ref{PExp}) as: 
\bea
\frac{V1}{M^3}&=& \frac{1}{4} (4 \lambda + \beta_\lambda),\nonumber \\
\frac{V2}{M^2}&=& \frac{1}{4} (12\lambda + 7\beta_\lambda+M \beta_\lambda^\prime), \nonumber \\
\frac{V3}{M}&=& \frac{1}{4} (24\lambda + 26\beta_\lambda+10M \beta_\lambda^\prime+M^2 \beta_\lambda^{\prime\prime}), 
\label{ICons2}
\eea
where the prime denotes $d/d\phi$.
Using $V1/M^3\simeq 0$ and $V2/M^2\simeq 0$, we obtain
\bea
 \beta_\lambda (M)\simeq -4\lambda(M), \qquad
 M\beta_\lambda^{\prime}(M)\simeq 16 \lambda (M). 
 \label{Cond1}
\eea
For small values of the couplings, $\lambda$, $g$, and $Y$,  we have $M^2 \beta_\lambda^{\prime\prime}(M) \simeq - M \beta_\lambda^{\prime}(M) \simeq -16 \lambda(M)$, where we have neglected higher order coupling terms such as $g^8$, $Y^8$, and $\lambda^4$. 
Hence the last equation in Eq.~(\ref{ICons2}) is simplified to $V3/M \simeq 16 \;\lambda(M)$. 
Comparing it with Eq.~(\ref{FEq-V3}), we obtain 
\bea
\lambda(M)\simeq 4.770 \times 10^{-16} \Big(\frac{M}{M_{P}}\Big)^2\Big(\frac{60}{N}\Big)^4,
\label{FEq1} 
\eea 
where we have approximated $V0\simeq (1/4) \lambda(M) M^4$. 
Since the $\lambda(M)$ is extremely small, we approximate $\beta_\lambda(M) \simeq 0$, which leads to 
\bea
Y(M)\simeq {32}^{1/4}\;g(M), 
\label{FEq3}
\eea
assuming that the beta-function is dominated by the gauge and the Yukawa couplings. 
This equation implies that the mass ratio between the right-handed neutrinos and the $B-L$ gauge boson is fixed to realize a successful inflection-point inflation. 
Using the second equation in Eq.~(\ref{Cond1}) and Eq.~(\ref{FEq3}), we find $\lambda(M)\simeq 3.713\times 10^{-3} \;g(M)^6$. 
Then from Eq.~(\ref{FEq1}), $g(M)$ is expressed as 
\bea
g(M)\simeq 7.107\times 10^{-3} \;\Big(\frac{M}{M_{P}}\Big)^{1/3}.
\label{FEq2} 
\eea
Finally, from Eqs.~(\ref{FEq-r}) and (\ref{FEq1}), the tensor-to-scalar ratio ($r$) is given by 
\bea
r \simeq 3.670 \times 10^{-9}  \Big(\frac{M}{M_{P}}\Big)^6, 
\label{FEqR} 
\eea
which is very small, as expected for the SFI scenario.

At the end of inflation $\epsilon(\phi_E)$ is explicitly given by
\bea
\epsilon(\phi_E) = \frac{{M_P}^2}{2 V0^2} \left(V1-V2 \;M \delta_E+\frac{V3}{2}M^2 {\delta_E}^2\right)^2 \simeq   \frac{{M_P}^2\;  M^6\;  {\delta_E}^2}{2\; V0^2 } \left(-\frac{V2}{M^2} +\frac{V3}{2 M} {\delta_E}\right)^2. 
\label{epsilon}
\eea  
We evaluate $\delta_E$ from $\epsilon(\phi_E)=1$. 
If we assume that the first term dominates in the parenthesis of the final expression above we find  $\delta_E\gg 1$ by using Eqs.~(\ref{FEq-V1V2}) and (\ref{FEq1}), which is inconsistent. 
Therefore, the second term dominates,  and hence we obtain
\bea
\delta_{E} \simeq 0.210 \; \Big(\frac{M}{M_{P}}\Big)^{1/2}, 
\label{delta}
\eea 
by using Eqs.~(\ref{FEq-V3}) and (\ref{FEq1}).

Before presenting our numerical analysis results, we check the consistency of our analysis. 
In our analysis in the previous section we have approximated the inflaton potential by Eq.~(\ref{PExp}), neglecting the higher order terms. 
For consistency, we need to check if the contribution of higher order terms can be neglected in our $B-L$ model. 
Consider the following expansion of inflaton potential at $\phi=M$, 
\bea
V(\phi) = \sum_{n=0} \frac{V^{(n)}}{n !} (\phi-M)^n, 
\label{PExpGen} 
\eea
where $V^{(n)}$ is the n-th derivative of the potential evaluated at $\phi = M$. 
As before, $V1=V^{(1)}$ and $V2=V^{(2)}$ are uniquely fixed by the  experimental values of scalar power-spectrum ($\Delta_\mathcal{R}^2$) and spectral index ($n_s$), respectively. 
For the consistency of our previous analysis, we require that the terms $V^{(4)}=V4$ and higher contribute negligibly in determination of $\delta_E$. 
Using Eqs.~(\ref{SRCond}) and (\ref{PExpGen}) $\epsilon(\phi_E)$ is expressed as
\bea
\epsilon(\phi_E) &\simeq& \frac{{M_P}^2}{2 V0^2} \left(\sum_{n=1} \frac{V^{(n)}}{(n-1)!} (\phi-M)^{n-1}\right)^2 \\ \nonumber
&\simeq& 
\frac{{M_P}^2}{2 V0^2} \left( \frac{V3}{2}M^2 {\delta_E}^2+ \sum_{n=4} \frac{V^{(n)}}{(n-1)!} (M \; \delta_E )^{n-1}\right)^2,  
\label{DelepsBound} 
\eea 
where we have used $V(\phi_E) \simeq V0$. 
This leads to constraint 
\bea
{\delta_E}^{(p-3)} <  \left|\frac{(p-1)!}{2} \frac{V^{(3)}}{V^{(p)}} M^{3-p}\right|. 
\label{DelBound} 
\eea
where $p\geq4$. 
To proceed further we need to evaluate Eq.~(\ref{DelBound}) explicitly for the minimal $B-L$ model.
As has been shown previously in this section, all the higher order derivatives of the potential can be approximately given by $V^{(n)} M^{n-4} \simeq C_n \lambda(M)$, where $C_n$ is a constant; for example, $C_4 = 96$ and $C_5=184$. 
We find that the most severe bound for both cases is from $V^{(4)}$ term.  
Using Eqs.~(\ref{delta}) we obtain upper bound on $M< 5.67 M_{P}$. 
%%%%%%%%%%%%%%%%%%%%%%%%%%%%%%%%%
\begin{figure}[h]
\begin{center}
\includegraphics[scale =1.45]{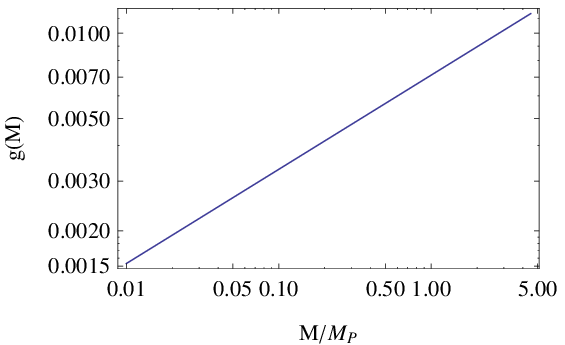} \;
\includegraphics[scale=1.45]{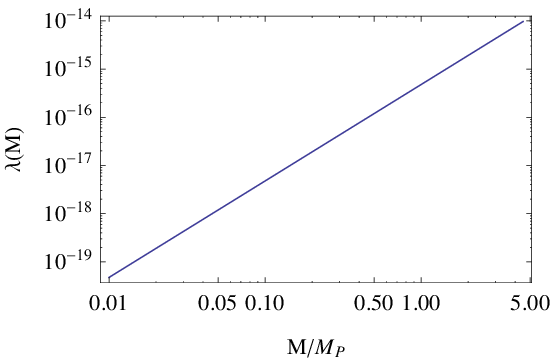}       
\end{center}
\caption{
Left panel and right panels show gauge coupling $g$ and inflaton quartic 
	coupling $\lambda$ values at  $M$ plotted against $M/M_{P}$, respectively. 
}
\label{fig1}
\end{figure}
%%%%%%%%%%%%%%%%%%%%%%%%%%%%%%%%%

Let us now present the numerical analysis of $B-L$ Higgs inflation scenario. 
For the rest of the paper, we set $e$-folding number $N=60$, and hence $M$ is the only free parameter in our analysis, and all the gauge, Yukawa and quartic coupling are expressed in terms of $M$. 
In Fig.~\ref{fig1}, we show the gauge coupling (left) and the quartic coupling (right) as a function of $M$. 
Imposing $M < 5.67 M_{P}$, we obtain an upper-bound on the couplings, 
\bea
g(M) < 1.261 \times 10^{-2}, \;\;\; Y(M) < 3.001 \times 10^{-2},\;\;\, \lambda(M) < 1.486 \times 10^{-14}.
\label{upperbound} 
\eea

%%%%%%%%%%%%%%%%%%%%%%%%%%%%%%%%
\begin{figure}[h]
\begin{center}
\includegraphics[scale =1.47]{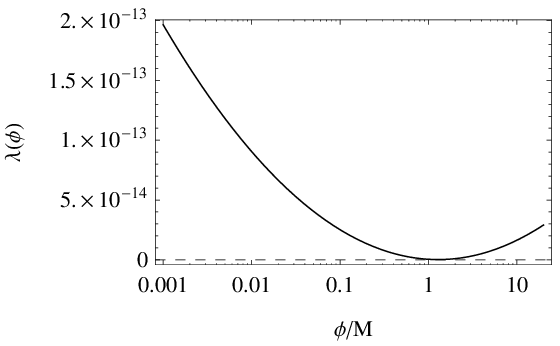} \;\;\;\;
\includegraphics[scale=1.42]{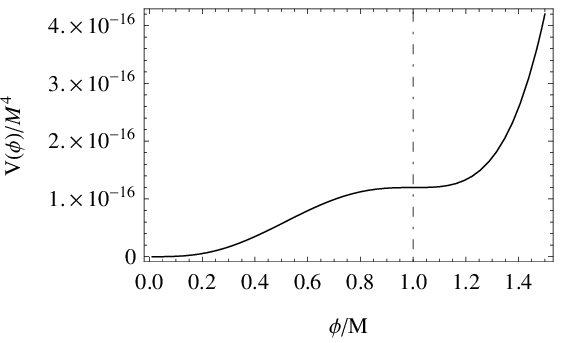}       
\end{center}
\caption{
Left panel shows the RG running of $B-L$ Higgs/inflaton quartic coupling plotted against the normalized energy scale $\phi/M$.
Here we have fixed $M = M_{P} $, so that  $g=7.107 \times 10^{-3}$, $Y= 1.690 \times 10^{-2}$, and $\lambda(M) \simeq 4.770 \times10^{-16}$. 
Dashed horizontal line corresponds to $\lambda=0$. 
Right panel shows the corresponding RG improved inflaton potential, 
	where the inflection-point-like point appears at $\phi=M$. 
}
\label{fig2}
\end{figure}
%%%%%%%%%%%%%%%%%%%%%%%%%%%%%%%%%%

In Fig.~\ref{fig2}, we plot the running quartic coupling (left) and the RG improved effective inflaton/Higgs potential (right). Here we have fixed $M=M_{P}$, which leads to $\lambda(M)\simeq 4.770 \times10^{-16}$, $g(M)\simeq7.107 \times 10^{-3}$ and $Y(M)\simeq1.690 \times 10^{-2}$. 
In the left panel, the dashed line corresponds to $\lambda=0$. 	
In the right panel, we see inflection-point-like behavior of the inflaton potential around $\phi=M$, marked with dashed-dotted line.

Now we discuss the inflationary predictions of our scenario. 
The prediction for the tensor-to-scalar ratio ($r$) is given by Eq.~(\ref{FEqR}). 
For the upper-bound on $M < 5.67 M_{P}$, the prediction for the tensor-to-scalar ratio is bounded by $r< 1.108\times10^{-4}$, which is very small. 
However, as discussed before the prediction for the running of the spectral index $\alpha \simeq - \; 2.742 \times 10^{-3}$, which is independent of $M$, is within the reach of future precision measurements ~\cite{RunningSpectral}.  
Note that this result is even independent of the particle content of the theory.

We now consider the particle mass spectrum of the model. 
At first, we evaluate the mass ratios of the right-handed neutrinos to the $Z^\prime $ gauge boson, $m_{N}/m_{Z^\prime}$, where $m_{N}$ is the degenerate right-handed neutrino mass.   
From the condition of almost vanishing $\beta_\lambda(M)$, we have $Y\simeq 32^{1/4} g$, and hence $m_{N}/m_{Z'}\simeq 0.84$ at M. 
This mass ratio remains almost the same at the $B-L$ Higgs VEV scale since the RG running effects for $g$ and $Y$ are negligible (see Eq.~(\ref{upperbound})).

As shown in Fig.~\ref{fig2}, although the inflection-point-like behavior requires $\beta_{\lambda}(M) \simeq - 4 \lambda(M) \simeq 0$, RG running  significantly changes the quartic coupling values at the low energies. 
In the following discussion we will derive an approximation formula for $\lambda$ at the low energies. 
Since $g(M)$, $Y(M)\ll 1$, the solutions to their RG equations are approximately given by 
\bea
g(\mu)  &\simeq& g(M) +\beta_{g}(M) \ln \left[\frac{\mu}{M}\right]  ,\nonumber \\ 
Y(\mu)  &\simeq& Y(M)+ \beta_{Y}(M) \ln \left[\frac{\mu}{M}\right], 
\label{gYatVEV}
\eea
where $\beta_{g}(M)$ and  $\beta_{Y}(M)$ are the beta-functions of $g$ and $Y$ at $M$, respectively (see Eq.~(\ref{RGEs})). 
Hence, the beta-function of the quartic coupling is approximately described as 
\bea
\beta_{\lambda}(\mu) &\simeq& 96 g^4(\mu) - 3 Y^4(\mu),  \nonumber \\ 
&\simeq& 4 \left\{ 96g(M)^3 \beta_{g}(M)-3Y(M)^3\beta_{Y}(M)\right\} \ln \left[\frac{\mu}{M}\right], \nonumber \\ 
&\simeq& 5.941 \times 10^{-2} g(M)^6 \ln \left[\frac{\mu}{M}\right], 
\label{betaapprox}
\eea 
where we have used $96g^4(M) \simeq 3 Y^4(M)$. 
Then we obtain the approximate solution to the RG equation as
\bea
\lambda(v_{BL})  &\simeq& \lambda(M) +3.868\times 10^{-15}\left(\frac{M}{M_P}\right)^2\left(\ln \left[\frac{v_{BL} }{M}\right] \right)^2, \nonumber \\
&\simeq & 3.868\times 10^{-15}\left(\frac{M}{M_P}\right)^2\left(\ln \left[\frac{v_{BL}}{M}\right] \right)^2, 
\label{lambdaatVEV}
\eea 
where $ v_{BL} \ll M$. 
Using $m_{Z^\prime} = 2 g(v_{BL}) v_{BL} \simeq 2 g(M) v_{BL} $, the mass ratio of the $B-L$ Higgs boson/inflaton to the $Z^{\prime}$ boson is found to be 
\bea
\frac{m_\phi}{m_{Z'}} \simeq 6.157 \times 10^{-6} \Big(\frac{M}{M_{P}}\Big)^{2/3} \ln \left[\frac{M}{v_{BL}}\right].
\label{ratiophiz} 
\eea

We now discuss a possibility of testing our scenario in the future collider experiments. 
In Ref.~\cite{ColliderConstraint}, the authors consider the heavy neutrino production at the High-Luminosity LHC \cite{HL-LHC} and the SHiP \cite{SHIP} experiments in the context of the minimal $B-L$ model. 
The process they have considered is a pair production of the heavy neutrinos via the decay of intermediate $Z^\prime$ boson, which is produced in proton-proton collisions. 
They focused on a simplified scenario where only one right-handed neutrino mixes with only one flavor of the SM neutrino via a small mixing angle. 
Since the lifetime of the heavy neutrino is long, 
its decay to the SM particles can be observed with a displaced vertex. 
For a fixed $m_{Z^{\prime}}/m_{N} = 3$, it has been found that the LHC and the SHiP experiments can explore the parameter regions,  $g \gtrsim 10^{-4}$ and $1 \lesssim m_{Z^{\prime}} [{\rm GeV}] \lesssim 500$.  
To implement this scenario, we extend our model with non-degenerate Majorana Yukawa couplings.  
For simplicity, we fix $Y_1$ to satisfy $m_{Z^{\prime}}/m_{N^1} = 3$, following Ref. ~\cite{ColliderConstraint}. 
We consider the remaining Yukawas to be degenerate, $Y_2= Y_3$. 
We repeat the same analysis as for the degenerate case, and find that $Y_{2, 3}\simeq 2.63 g $, or equivalently $m_{N^{2,3}}/m_{Z^\prime}\simeq 0.929$. 
We find that the constant coefficients in Eqs.~(\ref{FEq2}) and (\ref{betaapprox}) change to $7.120 \times 10^{-3}$ and $5.860 \times 10^{-2}$, respectively. 
These are coincidentally almost identical to those obtained in the degenerate case. 
As a result, the low energy value of the quartic coupling, and hence the mass ratio $m_\phi/m_{Z^{\prime}}$ is almost the same as that of the degenerate case. 
In the next section, we focus on this non-degenerate scenario.

%%%%%%%%%%%%%%%%%%%%%%%%%%%%%%%%%%%
\section{Constraints from the Big Bang Nucleosynthesis and the Current Collider Experiments}
%%%%%%%%%%%%%%%%%%%%%%%%%%%%%%%%%%%
Let us now consider a reheating scenario after the end of inflation  to connect our inflation scenario with the Standard Big Bang Cosmology. 
This occurs via inflaton decay into the SM particles while the inflaton oscillates around its potential minimum.   
We estimate the reheating temperature ($T_R$) as
\begin{eqnarray}
    T_R \simeq 0.55 \left(\frac{100}{g_*}\right)^{1/4} \sqrt{\Gamma M_P} .  
\label{TR}
\end{eqnarray} 
For the successful Big Bang Nucleosynthesis (BBN), we impose a model-independent  lower bound on 
   the reheating temperature as $T_R \gtrsim 1$ MeV.  
    
The general, renormalizable scalar potential for the $B-L$ Higgs/inflaton field and the SM Higgs doublet ($H$) is given by  
\bea
V(|H|,|\varphi|) =  \lambda \left(|\varphi|^2 -\frac{v^2_{BL}}{2}\right)^2 + \lambda_H \left(|H|^2 -\frac{v^2_H}{2}\right)^2 +  \lambda^{\prime} \left(|H|^2 -\frac{v^2_H}{2}\right)\left(|\varphi|^2 -\frac{v^2_{BL}}{2}\right), 
\label{InfPot}
\eea 
where $\lambda^{\prime}>0$ is the mixing term between the two scalar fields. 
The vacuum of the system is located at $\langle\varphi\rangle = v_{BL}/\sqrt{2}$ and $\langle H\rangle = (\frac{v_{H}}{\sqrt{2}} \;\;0)^T$. 
After the breaking of the $B-L$ and the  electroweak symmetries, the mass matrix is given by 
\begin{eqnarray}
{\cal L}  \supset -
\frac{1}{2}\begin{bmatrix}h  & \phi\end{bmatrix}
\begin{bmatrix} 
m_h^2 &  \lambda^{\prime} v_{BL} v_{H} \\ 
 \lambda^{\prime} v_{BL} v_{H} & m_{\phi}^2
\end{bmatrix} 
\begin{bmatrix} h \\ \phi \end{bmatrix}, 
\label{massmatrix}
\end{eqnarray} 
where $m_{\phi}^2 = 2 \lambda v_{BL}^2$, $h$ is the SM Higgs boson with mass $m_h = \sqrt{2 \lambda_{H}} v_{H} = 125$ GeV, where $\lambda_H$ is the SM Higgs quartic coupling and $v_{H}=246$ GeV. 
We diagonalize the mass matrix by 
\begin{eqnarray}
\begin{bmatrix} h \\ \phi \end{bmatrix}   =
\begin{bmatrix} \cos\theta &   \sin\theta \\ -\sin\theta & \cos\theta  \end{bmatrix} \begin{bmatrix} \phi_1 \\ \phi_2 
\end{bmatrix}  ,
\end{eqnarray} 
where $\phi_1$  and $\phi_2$ are the mass eigenstates. 
The relations among the mass parameters and the mixing angle ($\theta$) are the following: 
\bea
&2 v_{BL} v_{H}  \lambda^\prime= ( m_h^2 -m_\phi^2) \tan2\theta,   \nonumber  \\
 &m_{\phi_1}^2 = m_h^2     - \left(m_\phi^2  - m_h^2 \right) \frac{\sin^2\theta}{1-2 \sin^2\theta} , \nonumber \\
 &m_{\phi_2}^2 = m_\phi^2 + \left(m_\phi^2 - m_h^2 \right) \frac{\sin^2\theta}{1-2 \sin^2\theta} \ .
\label{mixings} 
\eea
Since the inflaton is much lighter than the $Z^\prime$ boson and the heavy neutrinos, it decays to the SM particles mainly through the mixing with the SM Higgs boson. 
We calculate the inflaton decay width as 
\bea 
   \Gamma_{\phi_2} = \sin^2\theta \times  \Gamma_h(m_{\phi_2}) ,
\eea
where $\Gamma_h(m_{\phi_2})$ is the SM Higgs boson decay width 
   if the SM Higgs boson mass were $m_{\phi_2}$.

There are constraints on the mixing angle. 
Firstly, the introduction of the mixing coupling modifies the beta-function of the inflaton quartic coupling in Eq.~(\ref{BGen}) as $16 \pi^2  \beta_\lambda \to 16 \pi^2 \beta_\lambda + 2 \lambda^{\prime 2}$.
In order not to change our results in the previous sections,  
    $\lambda^{\prime 2}$  should be negligibly small in the beta-function i.e 
 $\lambda^{\prime 2} \ll {48 g^4}$, evaluated at $M$. 
Another constraint on the mixing angle is from requiring positive definiteness of mass squared eigenvalues of the mass matrix in Eq.~(\ref{massmatrix}), which leads to $\lambda^{\prime 2} < 4 \lambda_{H} \lambda_{\phi}$. 
We find that the latter constraint is more severe and requires $\theta \ll 1$.  
Hence $\phi_1$ and $\phi_2$ are mostly the SM Higgs and the B-L Higgs mass eigenstates,  respectively.

In the following analysis we parameterize $\lambda^{\prime 2}  = 4 \lambda_{H} \lambda_{\phi}  \xi$, with a new parameter $0< \xi <1$. 
From Eq.~(\ref{mixings}), we obtain 
\bea
\theta^2 \simeq  \xi \left(\frac{m_{\phi}}{m_h}\right)^2,
\label{mixingangle}
\eea  
where we have used $m_\phi^2 \ll m_h^2$ for the parameter region we are interested in, namely 
 $1 \lesssim m_{Z^\prime}{\rm [GeV]} \lesssim 500$. 
We also find that $m_{\phi2} \simeq m_{\phi}\sqrt{1-\xi}$. 
From Eqs.~(\ref{ratiophiz}) and (\ref{mixingangle}), we can express the reheating temperature as a function of $M$,  $m_{Z^\prime}$ and $\xi$. 
For maximum value of $M = 4.6 M_{P}$ and a fixed $\xi$, there is a lower bound on the mass of $Z^{\prime}$ from the BBN constraint on reheating temperature, for example if $\xi = 0.5$, $0.1$, and $0.01$ we find $m_{Z^{\prime}} \gtrsim$ $13.6$, $21.5$, $46.5$ GeV, respectively. 
%%%%%%%%%%%%%%%%%%%%%%%%%%%%%%%%
\begin{figure}[h]
\begin{center}
\includegraphics[scale =1.65]{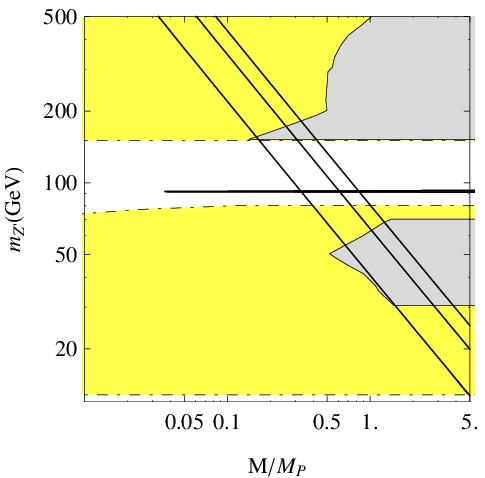}  \;\;\;
\includegraphics[scale =1.65]{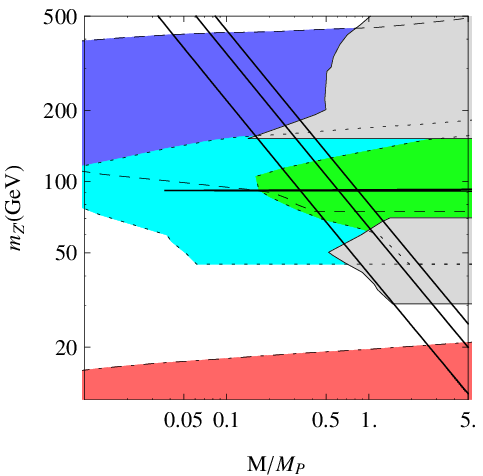}    

\end{center}
\caption{In both panels, the diagonal lines from left to right are the contours for the reheating temperature of $T_R =1$ MeV for fixed $\xi = 0.5$, $0.1$, and $0.05$, respectively.  
The regions to the left of the diagonal lines are excluded by the BBN constraint. 
For $m_{Z^\prime}/m_{N} = 3$, the (Grey) shaded regions bounded by the solid lines are excluded by the current LHC experiments. 
The very narrow regions bounded by the horizontal solid lines are excluded by the LEP experiment.  
The remaining regions bounded by broken lines are the future reach of different experiments, where the left (right) panel shows the future reach using direct (displaced vertex) searches.} 
\label{fig3}
\end{figure}
%%%%%%%%%%%%%%%%%%%%%%%%%%%%%%%%%%

Our results are shown in Fig.~\ref{fig3}.
In both panels, from left to right, the diagonal lines are contours corresponding to $T_R =1$ MeV for fixed $\xi = 0.5$, $0.1$ and $0.05$, respectively.  
The regions to the left of each contour are excluded by the BBN constraint.  
For the current experimental constraints and the future search reach we have referred to the results presented in Ref.~\cite{ColliderConstraint}.
Here, for fixed $m_{Z^{\prime}}/m_{N} = 3$, the authors have shown all the current experimental constraints and the future search reach on the parameter space of the minimal $B-L$ model, namely $m_{Z^{\prime}}$ and $g$. 
In Fig.~\ref{fig3}, we have re-parameterized the gauge coupling ($g$) in terms of $M/M_{P}$ according to Eq.~(\ref{FEq2}). 
In both of the panels, the excluded regions by the current LHC experiments are shown by the (Grey) shaded regions bounded by the solid lines.   
The very narrow regions bounded by the horizontal solid lines are excluded by the LEP experiment.  
In the left panel, the (yellow) shaded regions bounded by the dashed-dotted lines can be tested by the $Z^{\prime}$ resonance search at the High-Luminosity LHC. 
In the right panel, the regions bounded by the dashed and the dotted lines (dashed-dotted line) can be tested through the observation of the displaced vertex of the heavy neutrino decay at the High-Luminosity LHC (SHiP) experiment. 
For details about the correspondence of each shaded region to the individual experimental search, see Ref.~\cite{ColliderConstraint}. 
Combining the left and the right panels, all the allowed regions of our scenario for $13.6 \lesssim m_{Z^{\prime}}[{\rm GeV}] \lesssim500$ can be tested at the future experiments.

So far in our analysis, the inflaton is considered to be mostly the $B-L$ Higgs boson with a small mixing with the SM Higgs boson, which allows the inflaton to decay into the SM particles. 
Since the inflection-point-like inflation scenario requires $\lambda(v_{BL}) \ll 1$, we expect the reheating temperature to be small. 
We find that $T_{R} \lesssim 1$ GeV, for $m_{Z^{\prime}}\lesssim 500$ GeV. 
In terms of baryogenesis and dark matter physics, a reheating temperature $T_{R} \gg 1$ GeV is desirable.
Note that when we carefully consider the inflation trajectory on the scalar potential in Eq.~(\ref{InfPot}), the reheating process would be more involved. 
During the inflation, the inflaton tracks the trajectory with $H=0$. 
When the inflaton field rolls down to $\phi \simeq \sqrt{\lambda_H/\lambda^{\prime}} \; v_H$, the potential starts to develop a minimum away from $H=0$ in the $H$ direction. 
As the $\phi$ rolls down further, the minimum becomes deeper. 
Because the $\lambda \ll \lambda_H$, the potential is very flat along the $\phi$ direction at $H=0$, while it sharply drops in the $H$ direction at $\phi \simeq v_{BL}$.  
Hence, the SM Higgs field behaves like the waterfall field in the hybrid inflation scenario  \cite{HybridInflation}. 
In this case the decaying inflaton should have a sizable amount of the SM Higgs component. 
If the decaying inflaton field is mostly the SM Higgs field, the reheating temperature is evaluated by the decay width of the SM Higgs boson ($\Gamma_h\simeq 4.07 $ MeV), so that we find $T_{R} \simeq 10^7$ GeV. 
Although the true reheating temperature must be less than $10^7$ GeV, we may expect the actual reheating temperature sufficiently high, say, $T_R \gg 1$ TeV, which is desirable for the (thermal) baryogenesis and the dark matter physics.

If the reheating temperature is high, the $B-L$ Higgs boson $\phi$ can be in thermal equilibrium through its interactions with the SM Higgs boson and the heavy neutrinos. 
The interaction is so weak that $\phi$ decouples in the relativistic regime. 
We requires that $\phi$ decays before the BBN era, namely the lifetime of $\phi$ must be shorter than $1$ second ($\tau_\phi = 1/\Gamma_\phi < 1$ s), not to ruin the success of the BBN scenario. 
This condition is mathematically the same as requiring the reheating temperature $T_R> 1$ MeV from the inflaton $\phi$ decay. 
Hence, the results shown in the Fig.~\ref{fig3} are also applied to the scenario with a very high reheating temperature.

%%%%%%%%%%%%%%%%%
\section{Conclusion}
%%%%%%%%%%%%%%%%%
From a theoretical point of view, if the inflaton value is trans-Planckian, effective operators suppressed by the Planck mass could significantly affect the inflaton potential, and hence the inflationary predictions. 
To avoid this problem, we may consider slow-roll inflation with a small initial inflaton value ($\phi_I < M_{Pl}$). 
In this case, the inflection-point inflation is an interesting possibility to realize a successful slow-roll inflation
when inflation is driven by a single scalar field. 
To realize the inflection-point-like behavior
for the renormalization group (RG) improved effective $\lambda \phi^4$ potential, the running quartic coupling $\lambda(\phi)$ must exhibit a minimum
with an almost vanishing value in its RG evolution, namely $\lambda(\phi_I) \simeq 0$ and $\beta_{\lambda}(\phi_I) \simeq 0$, where $\beta_{\lambda}$ is the beta-function of the quartic coupling.

From a particle physics perspective, it is more compelling to consider an inflationary scenario, where the inflaton field plays another important role. 
We have considered a general Higgs model, with the gauge and the Yukawa interactions, and identified the Higgs field with the inflaton. 
In this case, the conditions, $\lambda(\phi_I) \simeq 0$ and $\beta_{\lambda}(\phi_I) \simeq 0$, lead to relations among the model parameters, the gauge, the Yukawa and the Higgs quartic couplings. 
Using the relations and requiring the inflationary predictions to be consistent with the Planck 2015 results \cite{Planck2015}, we have found that all the couplings depend only on $\phi_I$. 
Hence, the low energy mass spectrum of the model is uniquely determined by only two free parameters, $\phi_I$ and the inflaton/Higgs VEV. 
Hence the inflationary predictions are complimentary to the low energy mass spectrum. 
We have also shown that the inflection-point inflation provides a unique prediction for the running of the spectral index $\alpha \simeq  - 2.7 \times 10^{-3}\left(\frac{60}{N}\right)^2$, where $N$ is the $e$-folding number, independently of the model parameters. 
The future experiments can test this prediction for $\alpha$.

As an example Higgs model, we have  considered the minimal gauged $B-L$ extension of the Standard Model, and identified the $B-L$ Higgs field as the inflaton field. 
Hence, we have obtained our predictions for the mass spectrum for the $B-L$ gauge boson, the right-handed neutrinos, and the $B-L$ Higgs boson as a function of $\phi_I$ and the inflaton/Higgs VEV. 
We then considered the reheating after inflation. Imposing the Big Bang Nucleosynthesis constraint and the current collider experimental bounds, we have identified the allowed parameter regions. 
The entire parameter region for $m_{Z^\prime} < 500$ GeV can be tested by the future collider experiments such as the High-Luminosity LHC and the SHiP experiments.

%%%%%%%%%%%%%%%%%%%%%%%%%%%%%%%%%%%%%%%%%
\section*{Acknowledgements}
%%%%%%%%%%%%%%%%%%%%%%%%%%%%%%%%%%%%%%%%%
This work is supported in part by the United States Department of Energy (Award No.~DE-SC0013680).

%%%%%%%%%%%%%%

\end{document}